\documentstyle[12pt]{article}
\renewcommand
\baselinestretch{1.3}

\def\beq{\begin{equation}}
\def\brr{\begin{array}}
\def\err{\end{array}}
\def\eeq{\end{equation}}
\def\bea{\begin{eqnarray}}
\def\eea{\end{eqnarray}}
\def\bs{\bigskip}
\def\tr{\mbox{Tr}\, }
\def\ni{\noindent}
\def\wt{\widetilde}
\def\wh{\widehat}

\def\nn{\nonumber}
\def\ms{\medskip}

\def\dsp{\displaystyle}

\begin{document}

\hfill HUPD-92-19

\hfill December 1992

\vspace*{3mm}

\begin{center}

{\LARGE \bf
One-loop counterterms in 2d dilaton-Maxwell quantum gravity}

\vspace{4mm}

\renewcommand
\baselinestretch{0.8}
{\sc E. Elizalde}\footnote{E-mail address: eli @ ebubecm1.bitnet}
\\
{\it Department E.C.M., Faculty of Physics, University of
Barcelona, \\
Diagonal 647, 08028 Barcelona, Spain} \\
{\sc S. Naftulin}
\\
{\it Institut of Single Crystals, 60 Lenin Ave., 310141 Kharkov,
Ukraine} \\  and \\
{\sc S.D. Odintsov}\footnote{On sabbatical leave from
Tomsk Pedagogical Institute, 634041 Tomsk, Russia. E-mail address:
odintsov @ theo.phys.sci.hiroshima-u.ac.jp} \\ {\it Department of
Physics, Faculty of Science, Hiroshima University, \\
Higashi-Hiroshima 724, Japan}
\ms

\renewcommand
\baselinestretch{1.4}

\vspace{5mm}

{\bf Abstract}

\end{center}

The renormalization structure of two-dimensional quantum
gravity is investigated, in a covariant gauge. One-loop divergences
of the effective action are calculated. All the surface divergent
terms are taken into account, thus completing previous
one-loop calculations of the theory. It is shown that the on-shell
effective action contains only surface divergences. The off-shell
renormalizability of the theory is  discussed and  classes of
renormalizable dilaton and Maxwell potentials are found.



\newpage


The current activity in the study of two-dimensional (2d) quantum
gravity is motivated by the following (main) reasons. First of all,
it is extremely  difficult to study the physics of black holes and
of the early universe in the ambitious frames of the modern
models of four-dimensional quantum gravity. Solvable toy models
like 2d gravity can simplify the considerations and, hence, can
help in the study of more realistic situations.  From another
point of view, 2d quantum  gravity is interesting in itself, as it
is a good laboratory for developing the formal methods of quantum
field theory. Moreover, a connection with very fundamental physics
may be developed, because some models of 2d gravity are
string-inspired models.

Recently, an investigation of the renormalization structure of 2d
dilaton gravity has been started [1-4]. The one-loop counterterms
in covariant gauges have been calculated [1-4] and it has been
shown that the theory may be renormalizable off-shell for some
choices of the dilatonic potential (in particular, for the
Liouville potential). The one-loop divergences on shell are only
surface terms [1].

In the present paper we address the question of the calculation of
the one-loop counterterms which appear in dilaton-Maxwell 2d
quantum gravity as given by the following action\footnote{For a
higher derivative generalization see Ref. [12]}  [5,6]
\beq
S=-\int d^2x \, \sqrt{g}  \left[ \frac{1}{2} g^{\mu\nu}
\partial_{\mu} \varphi \partial_{\nu}  \varphi+ c_1R \varphi+
V(\varphi )  +\frac{1}{4} f
(\varphi)  g^{\mu\alpha}  g^{\nu\beta} F_{\mu\nu} F_{\alpha\beta}
\right],
\eeq
where  $F_{\mu\nu} =\partial_{\mu} A_{\nu}-\partial_{\nu}
A_{\mu}$ is the electromagnetic field-strength, $\varphi$ the
dilaton, $V(\varphi )$ is a general dilatonic potential and
$f(\varphi )$  is an arbitrary function. Notice that for some choices
of $f$ and $V$ this action corresponds to the heterotic string
effective action. The theory admits black hole solutions.
Properties of the 2d black holes associated with different variants
of the theory (1) have been studied in Refs. [5,9,10] along the same
lines that were developed for  pure dilaton gravity (without the Maxwell
term) [7,8]. The model given by (1) is also connected (via some
compactification) with the four-dimensional Einstein-Maxwell
theory,  which admits charged black hole solutions [14].

The one-loop counterterms for the theory with the action (1) have
been already calculated in Refs. [6,9], but {\it not} taking into
account the contributions of the divergent surface terms. Here
 we will generalize these calculations in order to take into
account all such terms in the efective action. It is well known
that they are relevant, e.g. in the context of Casimir effect
calculations (see, for example, Ref. [15]).

The classical field equations corresponding to (1) are
\bea
&& \nabla_{\mu} (f F^{\mu\nu}) =0, \ \ \ \ - \Box \varphi + c_1 R
+ V'(\varphi )+ \frac{1}{4} f'(\varphi )  F_{\mu\nu}^2 =0, \nn \\
&& -\frac{1}{2} (\nabla^{\alpha} \varphi) (\nabla^{\beta} \varphi)
+\frac{1}{4}g^{\alpha\beta} \nabla^{\mu} \varphi \nabla_{\mu}
\varphi +c_1 (\nabla^{\alpha} \nabla^{\beta}- g^{\alpha\beta}
\Delta) \varphi \\
&& +\frac{1}{2} g^{\alpha\beta} V +\frac{1}{8} g^{\alpha\beta} f
F_{\mu\nu}^2
 - \frac{1}{2} f F_{\ \mu}^{\alpha} F^{\beta\mu} =0. \nn
\eea
We will use Eqs. (2) later, in the analysis of the divergences of the
theory.

The background field method will be employed. The fields are split into
their quantum and background parts
\beq
\varphi \longrightarrow \bar{\varphi} = \phi +
\varphi, \ \ \  A_{\mu}  \longrightarrow \bar{A}_{\mu} =A_{\mu} +
Q_{\mu}, \ \ \
g_{\mu\nu} \longrightarrow \bar{g}_{\mu\nu} =g_{\mu\nu}
+h_{\mu\nu},
\eeq
where the second terms $ \varphi$, $Q_{\mu}$ and $h_{\mu\nu}$ are the
quantum fields. In what follows we shall use the dynamical
variables $h=g^{\mu\nu}h_{\mu\nu}$ and $\bar{h}_{\mu\nu}
=h_{\mu\nu}-\frac{1}{2}h g_{\mu\nu}$, rather than  $h_{\mu\nu}$.

In order to make contact with Ref. [4] ---where the surface divergences
in the absence of the Maxwell term have been calculated--- we add
to the action the following term
\beq
\Delta S=-\frac{1}{4} c_1 \, \xi \int d^2x \, \sqrt{g}  \left[
\phi \bar{h}_{\mu\nu} (R^{\mu\nu}-
\frac{1}{2} R g^{\mu\nu})
\right],
\eeq
where $\xi$ is an arbitrary parameter. Owing to the 2d identity
$R_{\mu\nu}- \frac{1}{2} R g_{\mu\nu}=0$, this expression (4) is
obviously zero. (Notice that $\Delta S$ is very similar to a  kind
of Wess-Zumino topological term. The fact that the parameter
$\gamma$ appears in the surface counterterms on-shell and here in
the anomaly, hints in this direction.)\, However, the second
variation of (4) may be important for the removal of non-minimal
contributions from the differential operator corresponding to
$S^{(2)}$. An interpretation of this term has been given in [4].

To fix the abelian and general covariant invariances, we use the
following covariant gauge conditions
\beq
 S_{GF}=\frac{1}{2}  \int d^2x \,   \left\{
\chi^A C_{AB} \chi^B\right\},
\eeq
where $A \equiv \{ \mu, * \}$, $\chi^* =- \nabla^{\nu} Q_{\nu}$, $
\chi^{\mu} = - \nabla^{\nu} \bar{h}^{\mu}_{\nu}- [c_1/(2\gamma
\phi)]  \nabla^{\mu} \varphi$, $\gamma = c_1 (\xi-1)/2$ and $C_{AB}
= \sqrt{g}$ diag $(f,2\gamma \phi g_{\mu\nu})$. If the identity (4)
is absent, then $\xi =0$ and $\gamma =-c_1/2$.

As is well known, the one-loop divergences of the one-loop
effective action are given by
\beq
\Gamma_{div} = \frac{i}{2} \tr \ln \wh{\cal H} - i \tr \ln \left.
\wh{\cal M}_{gh} \right|_{div},
\eeq
where  $\wh{\cal M}_{gh}$ is the ghost operator corresponding to
$S_{GF}$ (5) and the second variation of the action defines
$\wh{\cal H}$:
\beq
\frac{1}{2} \varphi^i \wh{\cal H}_{ij} \varphi^j = S^{(2)} + \Delta
S^{(2)} + S_{GF},
\eeq
being $ \varphi^i =( Q_{\mu}, \varphi, h, \bar{h}_{\mu\nu})$. From the
explicit forms of the action and background splittings (3), one
easily finds that
\beq
\wh{\cal H} =- \wh{K} \Delta +  \wh{L}^{\lambda} \nabla_{\lambda}
+ \wh{\cal M},
\eeq
where
\beq
\wh{K}^{-1} = \left( \brr{cccc}\dsp \frac{1}{f}
g_{\mu\alpha} & 0 & 0 & 0 \\ 0 & 0 & \dsp \frac{2}{c_1} & 0 \\ 0 &
\dsp \frac{2}{c_1} & \dsp - \left( \frac{4}{c_1^2}+ \frac{2}{\gamma
\phi} \right) & 0 \\ 0 & 0 & 0 & \dsp \frac{1}{\gamma
\phi} \wh{P}_{\mu\nu,\alpha\beta} \err
\right),
\eeq
being the projector $\wh{P}^{\mu\nu,\alpha\beta}  \equiv
\delta^{\mu\nu,\alpha\beta} -\frac{1}{2}
g^{\mu\nu} g^{\alpha\beta}$, and
\bea
\wh{L}^{\lambda}_{11} &=& f' (\nabla^{\alpha} \phi )
g^{\mu\lambda}- f' (\nabla^{\mu} \phi ) g^{\alpha\lambda}
- f' (\nabla^{\lambda} \phi ) g^{\mu\alpha}, \nn \\
\wh{L}^{\lambda}_{12} &=&-\wh{L}^{\lambda}_{21}=  f'
F^{\mu\lambda}, \ \ \ \ \wh{L}^{\lambda}_{13}
=-\wh{L}^{\lambda}_{31}= -\frac{1}{2} f
F^{\mu\lambda}, \nn \\
\wh{L}^{\lambda}_{14} &=&-\wh{L}^{\lambda}_{41}= f
F^{\lambda}_{\ \, \omega}\wh{P}^{\alpha\beta,\mu\omega} - f
F^{\mu}_{\ \, \omega}\wh{P}^{\alpha\beta,\lambda\omega}, \ \ \ \
\wh{L}^{\lambda}_{22}= \frac{c_1^2}{2\gamma\phi^2}
(\nabla^{\lambda} \phi ), \nn \\  \wh{L}^{\lambda}_{23} &
=&\wh{L}^{\lambda}_{32}= 0, \ \ \ \
\wh{L}^{\lambda}_{24} =-\wh{L}^{\lambda}_{42}=  (\nabla_{\omega}
\phi) \wh{P}^{\alpha\beta,\lambda\omega}, \ \ \ \
\wh{L}^{\lambda}_{33}=-\frac{c_1}{2}  (\nabla^{\lambda} \phi), \nn
\\
\wh{L}^{\lambda}_{34} &=&-\wh{L}^{\lambda}_{43}=
\frac{c_1}{2}
(\nabla_{\omega} \phi) \wh{P}^{\alpha\beta,\lambda\omega},  \\
\wh{L}^{\lambda}_{44} &=& \left( \frac{c_1}{2} -\gamma \right)
(\nabla^{\omega} \phi) \left(
\wh{P}^{\mu\nu}_{\omega\kappa} \wh{P}^{\alpha\beta,\lambda\kappa}-
 \wh{P}^{\mu\nu,\lambda\kappa} \wh{P}^{\alpha\beta}_{\omega\kappa}
\right) - \frac{3c_1}{2}
(\nabla^{\lambda} \phi) \wh{P}^{\mu\nu,\alpha\beta}, \nn  \\
\wh{M}_{11} &=& \frac{1}{2} fR\,g^{\mu\alpha}, \ \ \ \
\wh{M}_{23} =\wh{M}_{32}= \frac{1}{2} V'-
\frac{1}{8} f' F_{\mu\nu}^2, \ \ \ \ \wh{M}_{33}=
\frac{1}{8} f F_{\mu\nu}^2, \nn \\
\wh{M}_{44} &=& \left( \gamma -\frac{3}{2}\, c_1 \right)
(\nabla_{\lambda}
\nabla^{\omega}\phi ) \wh{P}^{\mu\nu,\lambda\kappa}
\wh{P}^{\alpha\beta}_{\omega\kappa} +  (\nabla_{\lambda} \phi) (
\nabla^{\omega}\phi ) \wh{P}^{\mu\nu,\lambda\kappa}
\wh{P}^{\alpha\beta}_{\omega\kappa} \nn \\
&-& \frac{1}{4}  (\nabla^{\lambda} \phi) ( \nabla_{\lambda}\phi )
\wh{P}^{\mu\nu,\alpha\beta} +\gamma   \phi R
\wh{P}^{\mu\nu,\alpha\beta} -\frac{1}{2}  V
\wh{P}^{\mu\nu,\alpha\beta} \nn \\
&-& \frac{1}{8}  f F_{\mu\nu}^2 \wh{P}^{\mu\nu,\alpha\beta} + f
F_{\omega\lambda}F^{\rho\lambda} \wh{P}^{\mu\nu,\omega\kappa}
\wh{P}^{\alpha\beta}_{\rho\kappa} +\frac{1}{2} f
F^{\omega\kappa}F^{\lambda\rho} \wh{P}^{\mu\nu}_{\omega\lambda}
\wh{P}^{\alpha\beta}_{\kappa\rho}. \nn
\eea

Now we can start the computation of the contribution of the
gravitational-Maxwell part to the effective action divergences.
First of all, we easily see that
\beq
  \tr \ln \left. \wh{\cal H}\right|_{div} =
\tr \ln \left. \left(-\wh{K}^{-1}  \wh{\cal H}\right)\right|_{div},
\eeq
because $\tr \ln (-\wh{K})$ gives a contribution proportional to
$\delta (0)$, which is zero in dimensional regularization. However,
in order to apply the standard algorithm
\bea
\Gamma_{div} &=& \frac{i}{2}  \tr \ln \left. \wh{\cal
H}\right|_{div} = \frac{i}{2}  \tr \ln \left(
 \wh{1} \Delta + 2 \wh{E}^{\lambda} \nabla_{\lambda}
+ \wh{\Pi} \right) \nn \\
&=& \frac{1}{2\varepsilon} \int d^2x \, \sqrt{g}\,  \tr \left(
\wh{\Pi} + \frac{R}{6} \, \wh{1} -  \nabla^{\lambda}
\wh{E}^{\lambda} - \wh{E}^{\lambda}\wh{E}_{\lambda} \right),
\eea
where $\varepsilon =2\pi (n-2)$ and where we should have
the operator $\wh{\cal
H}$  constrained to be hermitean, properly symmetrized, etc. The
explicit integrations by parts actually change the matrix elements
and lead to a breaking of these compulsory properties of $\wh{\cal
H}$.

In order to recover an operator  $\wh{\cal H}$ with the required
properties, the doubling procedure of 't Hooft and Veltman [13] is
very useful. A clear explanation on how to do this in the
present context can be found in Refs. [2-4] (for even more details
see [13]). In fact, using this method amounts to the following
redefinitions of the operator   $\wh{\cal H}$ as given in (8):
\bea
 \wh{L}_{\lambda} & \longrightarrow & \wh{L}_{\lambda}'=
\frac{1}{2} (\wh{L}_{\lambda} -\wh{L}_{\lambda}^T)
- \nabla_{\lambda} \wh{K}, \nn \\
\wh{\cal M} & \longrightarrow & \wh{\cal M}'=\frac{1}{2}(\wh{\cal
M}+ \wh{\cal
M}^T)-\frac{1}{2} \, \nabla^{\lambda} \wh{L}_{\lambda}^T -
\frac{1}{2} \, \Delta
\wh{K}.
\eea
Now, introducing  the notation $
\wh{E}^{\lambda} = -\frac{1}{2} \wh{K}^{-1} \wh{L}^{'\lambda}$ and
$ \wh{\Pi} = - \wh{K}^{-1} \wh{\cal M}'$, we have the operator
$\wh{\cal H}$ in the form (12) and we can apply the algorithm (12).

The explicit form of the operators $\wh{E}^{\lambda}$ and $
\wh{\Pi}$ can be found from (10) and (13) to be the following
\bea
(\wh{E}^{\lambda})^1_1 &=& \frac{f'}{2f} \left[
(\nabla^{\lambda} \phi) g^{\alpha}_{\rho}-  (\nabla^{\alpha} \phi)
g^{\lambda}_{\rho} +  (\nabla_{\rho} \phi) g^{\alpha\lambda}
\right], \ \ \ (\wh{E}^{\lambda})^1_2 = -\frac{f'}{2f} F^{\
\lambda}_{\rho}, \nn \\
(\wh{E}^{\lambda})^1_3 & =& \frac{1}{4} F^{\ \lambda}_{\rho}, \ \
\
(\wh{E}^{\lambda})^1_4 = \frac{1}{2} F_{\rho\omega}
\wh{P}^{\alpha\beta,\lambda\omega}- \frac{1}{2} F^{\lambda\omega}
\wh{P}^{\alpha\beta}_{\rho\omega}, \nn \\
(\wh{E}^{\lambda})^2_1 &= & -\frac{f}{2c_1} F^{\alpha \lambda}, \ \
\ (\wh{E}^{\lambda})^2_2 =( \wh{E}^{\lambda})^2_3 =0, \ \ \
(\wh{E}^{\lambda})^2_4 =-  \frac{1}{2} (\nabla_{\omega} \phi)
\wh{P}^{\alpha\beta,\lambda\omega}, \nn \\
(\wh{E}^{\lambda})^3_1 &= & \left( \frac{f}{c_1^2}+
\frac{f}{2\gamma\phi} + \frac{f'}{c_1} \right)
F^{\alpha\lambda},
\ \ \ (\wh{E}^{\lambda})^3_2 = - \frac{c_1}{2\gamma\phi^2}
(\nabla^{\lambda}
\phi), \ \ \ (\wh{E}^{\lambda})^3_3=0, \nn \\
(\wh{E}^{\lambda})^3_4& =&  \frac{c_1}{2\gamma\phi}
(\nabla_{\omega} \phi)
\wh{P}^{\alpha\beta,\lambda\omega}, \ \ \ (\wh{E}^{\lambda})^4_1=
\frac{f}{2\gamma\phi} \left( F^{\lambda}_{\ \omega}
\wh{P}^{\alpha\omega}_{\rho\sigma}-  F^{\alpha}_{\ \omega}
\wh{P}^{\lambda\omega}_{\rho\sigma}\right) , \nn \\
(\wh{E}^{\lambda})^4_2& =&  \frac{1}{2\gamma\phi} (\nabla_{\omega}
\phi) \wh{P}^{\lambda\omega}_{\rho\sigma}, \ \ \
(\wh{E}^{\lambda})^4_3=  \frac{c_1}{4\gamma\phi} (\nabla_{\omega}
\phi)
\wh{P}^{\lambda\omega}_{\rho\sigma}, \\
(\wh{E}^{\lambda})^4_4& =& \left( \frac{c_1}{4\gamma \phi}-
\frac{1}{2\phi} \right) (\nabla^{\omega} \phi)
\left( \wh{P}^{\alpha\beta}_{\omega\kappa}
\wh{P}^{\lambda\kappa}_{\rho\sigma} -
\wh{P}_{\rho\sigma,\omega\kappa}
\wh{P}^{\alpha\beta,\lambda\kappa} \right)+ \frac{1}{2\phi}
(\nabla^{\lambda} \phi) \wh{P}^{\alpha\beta}_{\rho\sigma}; \nn \\
\wh{\Pi}^1_1& =&-  \frac{1}{2} Rg^{\alpha}_{\rho}, \ \ \
\wh{\Pi}^2_2  = -  \frac{1}{c_1} V'+
\frac{f'}{4c_1}F_{\mu\nu}^2, \nn \\
\wh{\Pi}^3_3& =&-  \frac{1}{c_1} V' + \left(
\frac{f}{2c_1^2}+ \frac{f}{4\gamma\phi}+   \frac{f'}{4c_1}
\right)
F_{\mu\nu}^2 +\left(   \frac{1}{c_1}+ \frac{c_1}{2\gamma\phi}
\right)
(\Delta \phi), \nn \\
\wh{\Pi}^4_4& =&\left[ \left( \frac{3c_1}{2\gamma \phi}
-\frac{1}{\phi} \right)  \, (\nabla_{\lambda}
\nabla^{\omega} \phi)- \frac{1}{\gamma\phi}  \,
(\nabla_{\lambda}\phi)( \nabla^{\omega} \phi)
-\frac{f}{\gamma\phi}  \, F_{\lambda\nu}F^{\omega\nu} \right]
\wh{P}^{\lambda\kappa}_{\rho\sigma}
\wh{P}^{\alpha\beta}_{\omega\kappa}  \nn \\
&-& \frac{f}{2\gamma\phi}
F^{\omega\kappa} F^{\lambda\nu}
\wh{P}_{\rho\sigma,\omega\lambda} \wh{P}^{\alpha\beta}_{\kappa\nu}
\nn \\
&+& \left[ \frac{1}{4\gamma\phi}  \, (\nabla_{\lambda}\phi)(
\nabla^{\lambda} \phi)+ \left( \frac{1}{2\phi} -
\frac{3c_1}{4\gamma\phi} \right)
(\Delta \phi)- R + \frac{V}{2\gamma\phi}  +
\frac{fF_{\mu\nu}^2}{8\gamma\phi} \right]
\wh{P}^{\alpha\beta}_{\rho\sigma}.
\nn
\eea

Using the algorithm (12) and the operators (13), after some tedious
algebra we find the gravitation-Maxwell contribution to the
effective action, which is given by
\bea
\Gamma_{GM \ div}&=& -\frac{1}{2\varepsilon} \int d^2x \, \sqrt{g}
\left[
2R+ \frac{2}{c_1}\, V' - \frac{V}{\gamma\phi}+
 \left( \frac{f'}{2c_1} +
\frac{f}{4\gamma\phi} \right)F_{\mu\nu}^2\right. \\ &+& \left.
 \left( \frac{f'}{f}+ \frac{1}{\phi}-\frac{1}{c_1}
-\frac{c_1}{2\gamma \phi}
\right) (\Delta \phi) + \left( \frac{f''}{f}-
\frac{f^{'2}}{f^2}-\frac{1}{\phi^2}+\frac{c_1}{2\gamma \phi^2}
+\frac{c_1^2}{8\gamma^2 \phi^2} \right)
(\nabla^{\lambda} \phi) (\nabla_{\lambda} \phi) \right]. \nn
\eea

To complete the result we should find the ghost contribution to the
effective action divergences. The ghost operators corresponding to
(5) are
\beq
\wh{\cal M}^*_{gh} = \Delta, \ \ \ \ \wh{\cal M}^{\ \ \mu}_{gh \,
\nu} = g^{\mu}_{\ \nu} \Delta +\frac{c_1}{2\gamma \phi}
(\nabla_{\nu} \phi) \nabla^{\mu} +\frac{c_1}{2\gamma \phi}
(\nabla^{\mu} \nabla_{\nu} \phi) + R^{\mu}_{\ \nu}.
\eeq
Using again the algorithm (12), we get the full ghost contribution
to the effective action
\beq
\Gamma_{gh \ div}= -\frac{1}{2\varepsilon} \int d^2x \, \sqrt{g}
\left[
3R+\frac{c_1}{2\gamma \phi} (\Delta \phi )
+ \left( \frac{c_1}{2\gamma \phi^2} -\frac{c_1^2}{8\gamma^2 \phi^2}
\right)
(\nabla^{\lambda} \phi) (\nabla_{\lambda} \phi) \right].
\eeq
Finally, the total divergent part of the one-loop effective action
is given by the sum of (15) and (17)
\bea
\Gamma_{div}&=& -\frac{1}{2\varepsilon} \int d^2x \, \sqrt{g}
\left[
5R+ \frac{2}{c_1}\, V' - \frac{V}{\gamma\phi}+
 \left( \frac{f'}{2c_1} +
\frac{f}{4\gamma\phi} \right)F_{\mu\nu}^2\right. \nn \\ &+& \left.
 \left( \frac{f'}{f}+ \frac{1}{\phi}-\frac{1}{c_1} \right)
(\Delta \phi) + \left( \frac{f''}{f}-
\frac{f^{'2}}{f^2}-\frac{1}{\phi^2}+\frac{c_1}{\gamma \phi^2}
\right)
(\nabla^{\lambda} \phi) (\nabla_{\lambda} \phi) \right].
\eea

A few remarks are in order. First of all, in the absence of the
Maxwell sector ($f=0$) and dropping the surface divergent terms,
expression (17) (no surface divergences) coincides with the results
obtained in Refs. [2,3,4] in the same gauge
\beq
\Gamma_{div}= -\frac{1}{2\varepsilon} \int d^2x \, \sqrt{g} \left[
 \frac{2}{c_1}\, V' - \frac{V}{\gamma\phi}+ \frac{c_1}{\gamma
\phi^2} (\nabla^{\lambda} \phi) (\nabla_{\lambda} \phi) \right].
\eeq
Notice that in Refs. [2,3] the term (4) has not been introduced and
$\gamma =-c_1/2$. However, we actually disagree with Ref. [4] in
the surface divergent terms. This stems from the fact that the
term $-\wh{\nabla}^{\lambda} \wh{E}_{\lambda}$ corresponding to
the algorithm (12) is missing in Ref. [4]. This term gives in fact
an additional contribution to the surface divergent ones.

Second, dropping the surface divergent terms one can see that the
divergences of the Maxwell sector coincide with the ones previously
obtained in Refs. [6,9].

Eq. (18) ---which contains the surface divergent terms of the theory
(1)--- constitutes the main result of the present work, and completes
the calculations done in Refs. [6,9]. It is interesting to observe
that there is no dependence on $\gamma$ in the terms which result
from the Maxwell or Maxwell-dilaton sectors (only in the pure
dilaton sector does such dependence appear). Also to be remarked is
the fact that there is a contribution from the Maxwell terms to the
dilaton sector
\beq
 -\frac{1}{2\varepsilon} \int d^2x \, \sqrt{g} \, \nabla_{\lambda}
\left(  \frac{f'}{f} \, \nabla^{\lambda} \phi \right),
\eeq
which constitutes the surface term.

Let us now consider the result (18) on shell. Integrating (18) by
parts, keeping all the surface terms, and using the classical field
equations (2), we get the on-shell divergences of the effective
action
\beq
\Gamma_{div}^{on-shell}= -\frac{1}{2\varepsilon} \int d^2x \,
\sqrt{g} \left\{ 5R +
 \frac{2}{c_1} \left( V' + \frac{f'}{4} F_{\mu\nu}^2 \right)+
\Delta \left[ \ln f + \left( 1- \frac{c_1}{\gamma } \right) \ln
\phi - \frac{1}{c_1} \phi \right] \right\}.
\eeq
Using the second of the field equations (2), we can rewrite Eq.
(21) as
\beq
\Gamma_{div}^{on-shell}= -\frac{1}{2\varepsilon} \int d^2x \,
\sqrt{g} \left\{ 3R +
\Delta \left[ \ln f + \left( 1- \frac{c_1}{\gamma } \right) \ln
\phi + \frac{1}{c_1} \phi \right] \right\}.
\eeq
Hence, one can see that the one-loop divergences of the on-shell
effective action are just given by the surface terms. If we drop
the surface terms, we can say that the one-loop $S$-matrix is finite,
as in pure dilaton gravity [1]. Notice also that the surface
counterterms on shell depend on the parameter $\gamma$ (and, hence,
on $\xi$). This is not surprising because, by adding (4), one
introduces a new parameter into the theory. (This parameter
apparently can influence only surface counterterms, because of the
rather trivial structure of (4).)

Finally, let us discuss the renormalizability of the theory
off-shell. By dropping the surface terms in (18), we get
\beq
\Gamma_{div}= -\frac{1}{2\varepsilon} \int d^2x \, \sqrt{g} \left[
 \frac{2}{c_1}\, V' - \frac{V}{\gamma\phi}+
 \left( \frac{f'}{2c_1} +
\frac{f}{4\gamma\phi} \right)F_{\mu\nu}^2
+\frac{c_1}{\gamma \phi^2}
(\nabla^{\lambda} \phi) (\nabla_{\lambda} \phi) \right].
\eeq
For $\xi =0$ this has been obtained previously [9]. Adding to the
classical action (1) the counterterms ((23) with the opposite sign)
we obtain the renormalized action. Choosing the one-loop
renormalization of $g_{\mu\nu}$ as
\beq
g_{\mu\nu}= \exp \left( -\frac{1}{ 2\varepsilon \gamma \phi}
\right) \, \wt{g}_{\mu\nu},
\eeq
we get the renormalized action in the following form
\beq
S_R =  \int d^2x \, \sqrt{\wt{g}} \left[
 \frac{1}{2}  \wt{g}^{\mu\nu}
\partial_{\mu} \phi \partial_{\nu}  \phi+ c_1 \wt{R} \phi+
V(\phi )  +\frac{1}{4} f
(\phi)  F_{\mu\nu}^2 -\frac{V'(\phi )}{\varepsilon c_1}-
\frac{f'(\phi )}{4\varepsilon c_1}  F_{\mu\nu}^2
\right].
\eeq
The dilaton and the coupling $c_1$ do not get renormalized in the
one-loop approximation.

{}From (25) it follows that the theory under discussion is one-loop
multiplicatively renormalizable for the families of potentials:
\bea
V(\phi )=e^{\alpha\phi} + \Lambda, & & f(\phi )=e^{\beta\phi} \ \
\mbox{or} \ \   f(\phi )=f_1, \nn \\
V(\phi )=A_1 \sin \phi + B_1 \cos \phi, & & f(\phi )=A_2 \sin \phi
+ B_2 \cos \phi,
\eea
where $\alpha$, $\Lambda$, $\beta$, $f_1$, $A_1$,  $B_1$, $A_2$ and
$B_2$ are arbitrary constants.

The black hole solutions which appear for the Liouville like
potentials of (26) have been discussed in Refs. [5,9,10]. It would be
interesting to study the time-dependent solutions (2d cosmology)
for the theory that has been considered here with the potentials
(26) and taking into account the back reaction [11].

Summing up, we have investigated the renormalization structure of
Maxwell-dilaton gravity and found all the one-loop counterterms,
including the surface counterterms. The on-shell limit of the
effective action and the renormalizability off-shell have also been
studied in detail.

\vspace{5mm}

\ni{\large \bf Acknowledgments}

S.D.O. wishes to thank the Japan Society for the Promotion of
Science (JSPS, Japan) for finantial support and the Particle
Physics Group at Hiroshima University for kind hospitality.
E.E. has been supported by DGICYT (Spain), research project
PB90-0022, and by the Alexander von Humboldt Foundation (Germany).
\bs

\newpage

\renewcommand
\baselinestretch{1.2}

\end{document}